\documentclass[prb,aps,twocolumn,showpacs]{revtex4-1}
\usepackage{amsfonts,amsmath,bm}
\usepackage[OT4]{fontenc}
\usepackage{mathtools}
\usepackage{graphicx}
\usepackage{epstopdf}
\usepackage{color}

\newcommand{\rr}{{\bm{r}}}

\newcommand{\kp}{\mbox{$\bm{k}\!\cdot\!\bm{p}$ }}

\newcommand{\sumq}{\sum_{\bm{q}}}

\newcommand{\kk}{\bm{k}}
\newcommand{\qq}{\bm{q}}

\newcommand{\ket}{\rangle}

\newcommand{\rl}{\rangle\!\langle}

\newcommand{\ff}{\mathcal{F}}

\DeclarePairedDelimiter\norm{\lVert}{\rVert}%
\DeclarePairedDelimiter\abs{\lvert}{\rvert}%
\makeatletter
\let\oldabs\abs
\def\abs{\@ifstar{\oldabs}{\oldabs*}}
\let\oldnorm\norm
\def\norm{\@ifstar{\oldnorm}{\oldnorm*}}
\makeatother
\begin{document}
	
\title{Polaron resonances in two vertically stacked quantum dots}  
\author{Pawe{\l} Karwat}
\email{Pawel.Karwat@pwr.edu.pl}
\author{Krzysztof Gawarecki} 
\author{Pawe{\l} Machnikowski} 
\affiliation{Department of Theoretical Physics, Faculty of Fundamental Problems of
         Technology, Wroc{\l}aw University of
	Science and Technology, 50-370 Wroc{\l}aw, Poland}
	
\begin{abstract}
In this work, we present a theoretical study of polaron states in a double quantum dot
system. We present realistic calculations which combine 8 band \kp model, configuration
interaction approach and collective modes method.  We investigate the dependence of
polaron energy branches on axial electric field. We show that coupling between carriers
and longitudinal optical phonons via Fr\"{o}hlich interaction leads to qualitative and
quantitative reconstruction of the optical spectra. In particular, we study the structure
of resonances between the states localized in different dots. We show that $p$-shell
states are strongly coupled to the phonon replicas of $s$-shell states, in contrast to the
weak direct $s$-$p$ coupling. We discuss also the dependence of the phonon-assisted tunnel
coupling strength on the separation between the dots.
\end{abstract}
	
	\pacs{
		78.67.Hc, 
		71.38.-k, 
	}
	
\maketitle
\section{Introduction}
\label{sec:intro}
	
Self assembled quantum dots (QDs) are continuously attracting attention both in
fundamental research, as 
well as in the development of novel applications in quantum optics and quantum
information. With the continuing effort 
and progress in miniaturization, QDs find technological use in many types of devices,
including QD lasers\cite{zhuo14}, screens, solar cells\cite{schaller04} and many
others. 

One of the most interesting aspects of QD physics is related to carrier-phonon
coupling. Apart from dissipative processes induced by phonons, this coupling can lead to
the formation of polarons, that is, eigenstates of the interacting carrier-phonon systems
in which the carrier state is correlated with the coherent field of longitudinal optical
(LO) phonons. In QDs, the carrier spectrum is discrete, while the relatively weak carrier
localization limits the effectively coupled LO phonons to the nearly dispersionless
zone-center part of their spectrum. As a result, the system is in the strong coupling
regime and the polaron states are manifested in the form of pronounced resonances whenever
one excited state spectrally crosses a LO-phonon replica of another state
\cite{ham99,ham02,kaczmarkiewicz10}.  The width of the resonances provides a natural
quantitative measure of the strength of the carrier-phonon coupling. 
The effectively dispersionless nature of LO phonons forming the
polaron states in QDs makes it possible to describe the system in terms of a finite number
of collective modes \cite{stauber00}, which opens the path to numerically exact
diagonalization of the carrier-LO-phonon (Fr\"ohlich) Hamiltonian in a restricted basis
of carier states. Experimental and theoretical work on
QD polarons has brought good 
understanding of their essential properties both for single-electron states and for excitons
\cite{verzelen02a}, as well as of their crucial role for carrier relaxation in
self-assembled QD systems, where typical energy level separations are comparable to the LO
phonon energy \cite{verzelen00,jacak02a,verzelen02b,zibik04,zibik09}.

Systems composed of vertically stacked coupled QDs offer reacher physical properties and a
higher level of 
controllability than a single QD. In particular, a double quantum dot (DQD) structure
supports spatially direct and indirect states with different dipole moments, the energy of
which can be tuned in a broad 
range by applying an axial electric field
\cite{szafran05,szafran08,krenner05b,bracker05,muller12}. Recently, the 
spectrum of such a system was 
mapped out by combined spectroscopy techniques and successfully modeled using an 8-band
\kp theory in the envelope-function approximation
\cite{ardelt16}. The electric-field tunability of energy levels in such systems might allow
one to study the polaron resonances as a function of the electric field by matching
various energy shells of the two dots, which offers much more flexibility in comparison to
the single-QD studies, where only limited tunability by magnetic field is available \cite{ham99,ham02}.
	
In this paper we study polaron states in a DQD structure. In such a coupled structure, a
prerequisite of any quantitatively reasonable modeling is an accurate model of wave
functions. Therefore, in order to find electron and hole states we apply the 8-band
\kp model with strain distribution found within continuous elasticity
approach\cite{pryor98b}. We then calculate exciton states using the configuration interaction
method. Finally, polaron states are found by orthogonalization of the Fr\"ohlich
Hamiltonian in the basis of collective phonon modes\cite{stauber00}. We propose a
numerically efficient scheme of mode orthogonalization and selection of effectively
coupled modes.  We study the system
spectrum, focusing on the polaron resonances, i.e., the spectral anti-crossing structures
appearing when the energies of two carrier states differ by one LO phonon energy. We show
that the width of such a LO-phonon-assisted resonance between direct and indirect exciton
states of the same symmetry follows an exponential dependence on the inter-dot separation
with a similar exponent but lower amplitude, as compared to the direct resonance. In
contrast, for a pair of states with different symmetry, where the direct resonance is only
allowed by weak spin-orbit effects, the coupling mediated by LO phonons is much stronger
than the direct one. 
	
The paper is organized as follows. In Sec~\ref{sec:model}, we define the system under
consideration and the theoretical model. Next, in Sec.~\ref{sec:results} we present the
results. Section \ref{sec:conclusions} contains the final discussion. In the Appendix we describe the mode orthogonalization method.

\section{Model and numerical method}
\label{sec:model}

In this section we first describe the carrier system and its model used in our calculations and
summarize the essential features of the exciton spectrum to facilitate further discussion
of the polaronic effects (Sec.~\ref{sec:carriers}). Next, we define the LO-phonon-related
part of the model  (Sec.~\ref{sec:phonons}).
Then, we present the collective mode method with the mode
orthogonalization scheme (Sec.~\ref{sec:collective}).

\subsection{Carrier system and its model}
\label{sec:carriers}

\begin{figure}[t]
		\begin{center}
			\includegraphics[width=6cm]{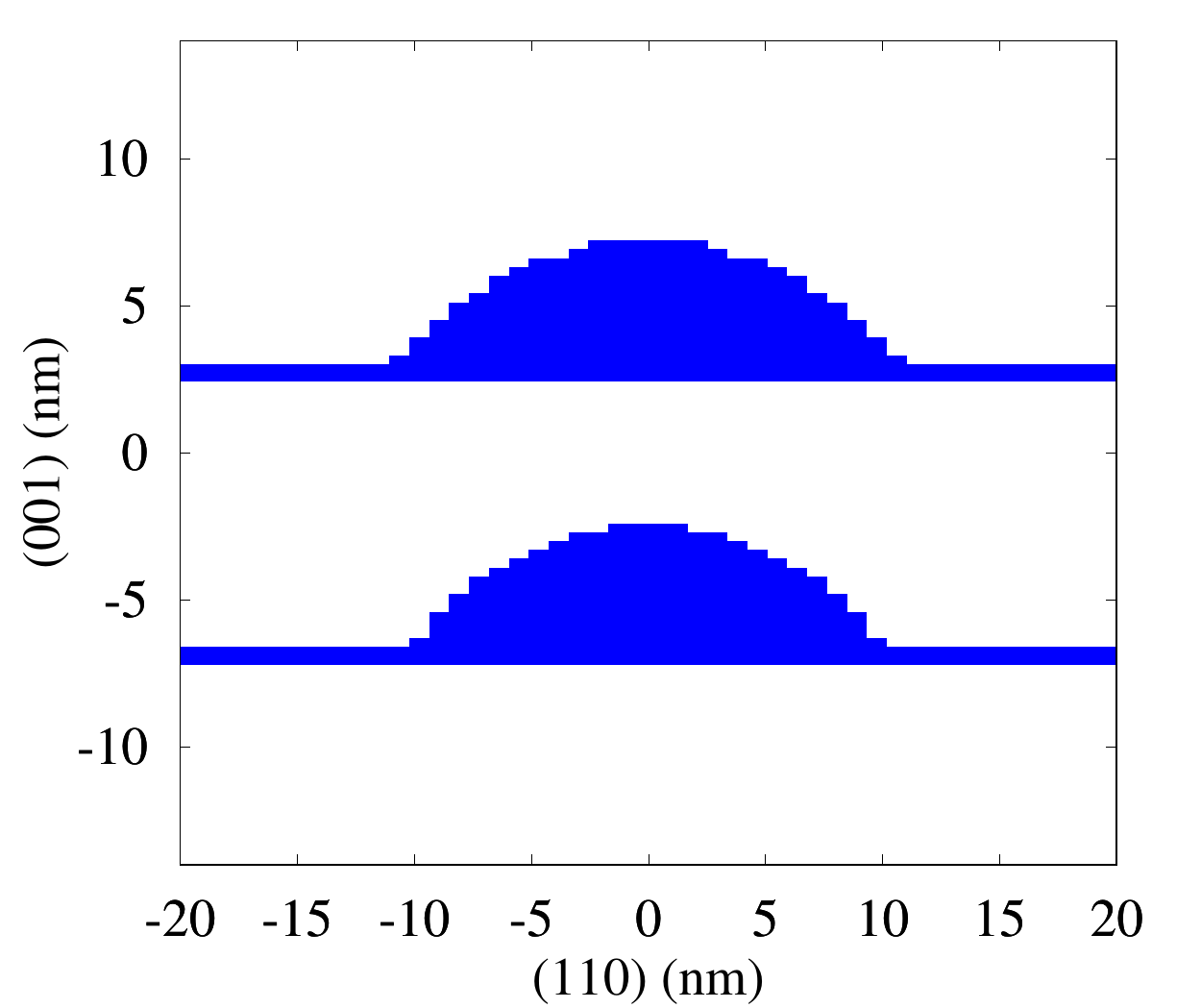}
		\end{center}
		\caption{\label{fig:composition} Material distribution in the system.}
\end{figure} The system under study is made up of two vertically stacked
InGaAs/GaAs self-assembled QDs resting on wetting layers. We assume lens shape of the upper (u) and lower (l) dot. In our calculations we take height $h = 4.2$~nm (the
same for both dots) and base radii $r_{\mathrm{l}} = 10.2$~nm, $r_{\mathrm{u}} = 10.8$~nm. The local InAs content in each QD and in the wetting layers is $80$\% InAs, the matrix contains $100$\% GaAs. The cross section 
of the InGaAs distribution is shown in Fig.~\ref{fig:composition}.
	
The Hamiltonian of a system of carriers coupled to LO phonons is
\begin{equation*}
H=H_{\mathrm{0}} + V_{\mathrm{c}} +  V_{\mathrm{ef}} + H_{\mathrm{ph}}+H_{\mathrm{F}},
\end{equation*}
where $H_{\mathrm{0}}$ describes the single particle states, $V_{\mathrm{c}}$ is
the~Coulomb interaction between the particles, $V_{\mathrm{ef}}$ represents an axial electric
field, $H_{\mathrm{ph}}$ is the~Hamiltonian of the LO phonon bath and $H_{\mathrm{F}}$
is a~Fr\"{o}hlich interaction between carriers and LO phonons. 

The first term of the Hamiltonian is 
\begin{equation*}
H_{\mathrm{0}} = \sum_{n}\epsilon^{(\mathit{e})}_{n} a^{\dagger}_{n} a_{n} 
+ \sum_{m} \epsilon^{(\mathit{h})}_{m} h^{\dagger}_{m} h_{m},
\end{equation*}
where $\epsilon^{(\mathit{e})}_{n}$/$\epsilon^{(\mathit{h})}_{m}$ are the energies of the
electron/hole states obtained from the $8$ band \kp calculations,
$a^{\dagger}_{n}$/$h^{\dagger}_{m}$, $ a_{n}$/$ h_{m}$ are the creation and
annihilation operators of the electron/hole in the state $n/m$, respectively. The details of the model are described in Ref.~\onlinecite{gawarecki14}.

The Coulomb
interaction is
\begin{equation*}
V_{\mathrm{c}} = \sum_{n n' \\ m m'} v_{n m m' n'} a^{\dagger}_{n} h^{\dagger}_{m} h_{m'} a_{n'},
\end{equation*}
where 
\begin{align*}
v_{n m m' n'} =& -\frac{e^{2}}{4 \pi \epsilon_{0} \epsilon_{\infty}} \int d \rr 
\int d \rr' \Psi^{*(\mathit{e})}_{n} (\rr) \Psi^{*(\mathit{h})}_{m} (\rr')   \\ 
& \times \frac{1}{\vert \rr-\rr' \vert} \Psi^{(\mathit{h})}_{m'} (\rr') \Psi^{(\mathit{e})}_{n'} (\rr).
\end{align*}
Here $e$ is the electron charge, $\epsilon_{0}$ and $\epsilon_{\infty}$ are vacuum
permittivity and high frequency dielectric constant, $\Psi^{(\mathit{e})}_{n}
(\rr)$/$\Psi^{(\mathit{h})}_{m} (\rr')$ corresponds to the electron/hole wave functions
which, according to our \kp model, are $8$ components spinors. For the sake of efficiency,
we calculate $v_{n m m' n'}$ in the inverse space (see details in 
Ref.~\onlinecite{daniels13}). 

The potential of an axial electric field is defined by
\begin{equation*}
V_{\mathrm{ef}} = \sum_{nn'}Z^{(\mathit{e})}_{nn'} a^{\dagger}_{n} a_{n'} 
- \sum_{mm'} Z^{(\mathit{h})}_{mm'} h^{\dagger}_{m} h_{m'},
\end{equation*}
where
\begin{equation*}
Z^{(\textit{e}/\textit{h})}_{ij} = 
F_{z} \int d \rr \Psi^{*(\textit{e}/\textit{h})}_{i}(\rr) 
z \Psi^{(\textit{e}/\textit{h})}_{j} (\rr)
\end{equation*}
and $F_{z}$ is the magnitude of the electric field and $H_{z}$ is the size
of our computational domain in the $z$ direction. 

\begin{figure}
\begin{center}
\includegraphics[width=230pt]{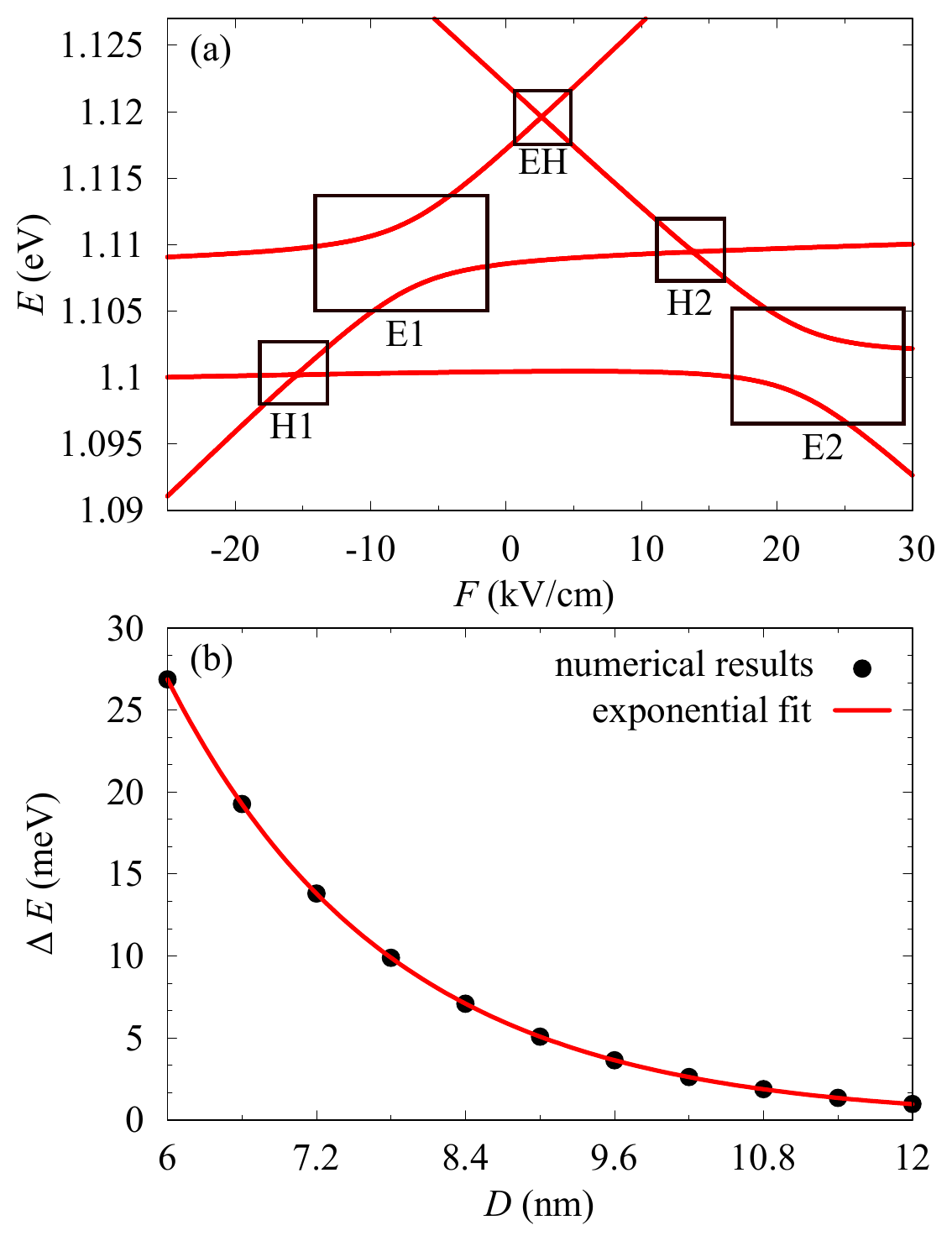}
\end{center}
\caption{\label{fig:x_res}(a) Exciton energy branches as a function of external axial electric field. (b) The width of the anticrossing E2 as a function of the inter-dot distance.}
\end{figure}
In our calculations, we first find the single-particle states using the \kp model. The strain distribution in the system is taken into account within the standard continuous elasticity framework\cite{pryor98b}. Piezoelectricity is included up to the second
order in strain tensor elements\cite{bester06} using parameters from
Ref.~\onlinecite{caro15}. In order to find electron and hole states, we perform the
calculation using the 8-band \kp method in the envelope function
approximation. The calculation details have been widely described in
Refs.~\onlinecite{gawarecki14,andrzejewski10} and material parameters are taken from Ref. \onlinecite{vurgraftman03}.
Having the single-particle states we then compute the exciton states which are found using the configuration interaction (CI) method. The axial field is included at the CI stage \cite{swiderski16}. Due to numerical efficiency reasons we limit
our basis to $4$ lowest electron and hole 
states (i.e. electron and hole $s$-shell in the lower and in the upper dot). 
The exciton spectrum obtained in this way\footnote{Since the
	LO-phonon coupling is not included explicitly in this calculation,  we have 
	replaced $\epsilon_{\infty}$ by $\epsilon_{s}$ in $V_{\mathrm{c}}$ to account for the
	LO-phonon induced contribution to the screening.} is presented in Fig.~\ref{fig:x_res}(a). 
This well-known spectrum of excitons in an electric field \cite{szafran05,szafran08}
is composed of spatially direct and indirect exciton states, clearly
distinguishable by the small and large slope of the field dependence of their energies, respectively.
If two such states are tuned into the resonance (and if the symmetry of the states is such
that selection rules are met) avoided crossing appears in the energy
spectrum\cite{szafran05,szafran08,bracker05,krenner05b,muller12}. In Fig.~\ref{fig:x_res}(a)
these resonance structures are marked by E1,E2 (electron tunneling resonance), H1,H2 (hole
tunneling) and EH (very weak coupling between two indirect configurations). We note that
extending the basis would lead to a more 
complex pattern including Coulomb resonances \cite{daniels13,ardelt16}.
Fig.~\ref{fig:x_res}(b) shows the electron avoided crossing width (E2) as a function of the distance $D$ between the dots. The dependence follows the exponential law\cite{bayer02b}. We obtained an excellent fit for $f(D) = a \exp(-b D)$ with $a = 0.751$~eV and $b = 0.555$~eV/nm. 
	 
\subsection{LO phonons}
\label{sec:phonons}

In the polaron formation we assume non-dispersive LO phonon modes with 
$\hbar\Omega = 36$~meV.
The LO phonon bath is then described by the Hamiltonian
\begin{equation*}
H_{\mathrm{ph}} = \sum_{\bm{q}}\hbar\Omega b^{\dagger}_{\bm{q}}b_{\bm{q}},
\end{equation*}
where $b^{\dagger}_{\bm{q}}$ and $b_{\bm{q}}$ are, respectively, the creation and
annihilation operators for the LO phonon mode $\bm{q}$. 
The carrier-phonon coupling is modeled by the Fr\"{o}hlich Hamiltonian,
\begin{align*}
H_{\mathrm{F}}=&-\sum_{\bm{q}}\frac{e}{q}\sqrt{\frac{\hbar \Omega}{2V\widetilde{\epsilon}\epsilon_{0}}}
\left ( \sum_{n n'} \mathcal{F}^{(\mathit{e})}_{n n'}(\qq) a^{\dagger}_{n} a_{n'} \right . \nonumber \\
& \left . - \sum_{m m'} \mathcal{F}^{(\mathit{h})}_{m m'}(\qq) h^{\dagger}_{m} h_{m'}
\right) \left( b^{\phantom{\dagger}}_{\bm{q}} + b^{\dagger}_{\bm{-q}}   \right ),
\end{align*} 
where $\widetilde{\epsilon}=(1/\epsilon_{\infty}-1/\epsilon_{s})^{-1}$, $\epsilon_{s}$ is
a static dielectric constant in GaAs, $V$ is the normalization volume for phonon modes and
$\mathcal{F}^{(\textit{e}/\textit{h})}_{nn'}(\qq)=\mathcal{F}^{(\textit{e}/\textit{h})*}_{n'n}(-\qq)$ 
is the one-particle (electron or hole) form-factor,
\begin{equation*}
\mathcal{F}^{(\textit{e}/\textit{h})}_{ij}(\qq) = 
\int \Psi^{*(\textit{e}/\textit{h})}_{i}(\rr) \Psi^{(\textit{e}/\textit{h})}_{j}(\rr) e^{i \kk \cdot\rr} d\rr.
\end{equation*}

We perform calculations for excitonic polaron states in the basis of noninteracting
electron and hole configurations  
$|\nu \ket  = |n\ket_{\mathrm{el}} \otimes |m\ket_{\mathrm{h}}$.  In this pair-state basis
the Fr\"ohlich Hamiltonian for two-particle (exciton) states has the form 
\begin{equation}\label{HF-X}
H_{\mathrm{F}}=-\sum_{\bm{q}}\frac{e}{q}\sqrt{\frac{\hbar \Omega}{2V\widetilde{\epsilon}\epsilon_{0}}}
\sum_{\nu\nu'} \mathcal{F}^{(\mathit{X})}_{\nu\nu'}(\qq)  |\nu\rl\nu'|
\left( b^{\phantom{\dagger}}_{\bm{q}} + b^{\dagger}_{\bm{-q}}   \right ),
\end{equation} 
where
$\mathcal{F}^{(\mathrm{x})}_{\nu\nu'} = 
\mathcal{F}^{(\mathit{e})}_{n n'}(\qq) \delta_{m  m'} - \mathcal{F}^{(\mathit{h})}_{m
  m'}(\qq) \delta_{n n'}$ with $\nu\sim (nm)$, $\nu'\sim(n'm')$.

\subsection{Collective modes}
\label{sec:collective}

The direct diagonalization of the Hamiltonian would imply sampling the $\qq$ space, which
is not feasible due to the
large number of required  $\qq$ points. However, for 
non-dispersive LO phonon modes, one can use the collective modes
method \cite{stauber00}. Thus, one defines the collective modes corresponding to the annihilation operators
\begin{displaymath}
\widetilde{B}_{\nu \mu}=
\sum_{\qq}\sqrt{\frac{l_{0}}{V}} \frac{1}{q} \mathcal{F}^{(\mathrm{x})}_{\nu\mu} b^{\phantom{\dagger}}_{\bm{q}},
\end{displaymath}
where $l_{0}$ is an arbitrary characteristic length. 
The Fr\"ohlich Hamiltonian then becomes
\begin{equation}\label{HF-coll}
H_{\mathrm{F}}=\sum_{\nu \mu}\sqrt{\frac{\hbar \Omega e^{2}}{2 l_{0} \widetilde{\epsilon}\epsilon_{0}}}
|\nu\rl\mu|\widetilde{B}_{\nu \mu} + \mathrm{h.c.}
\end{equation}  
However, the collective phonon modes 
$\widetilde{B}_{\nu \mu}$ are not orthonormal in the sense of canonical commutation
relations. Indeed, their commutator is
\begin{equation}
\label{gram_matrix}
A_{(\nu' \mu')(\nu \mu)} =
[\widetilde{B}_{\nu \mu},\widetilde{B}^{\dagger}_{\nu' \mu'}] 
= \frac{l_{0}}{V} \sum_{\qq} \frac{1}{q^{2}} \mathcal{F}^{*}_{\nu' \mu'} (\qq) \mathcal{F}_{\nu \mu} (\qq)
\end{equation}
and is related to the orthogonality of the form-factors, which does not hold in general.

At this point one has to choose between the original approach
involving orthogonalization of the modes \cite{stauber00} and the alternative method of
non-orthogonal modes \cite{obreschkow07}. The latter saves one computational step required
for orthogonalization but becomes inconvenient when the form-factors defining the modes
are not guaranteed to be 
linearly independent, which is the case, e.g., for the Fock-Darwin model
\cite{kaczmarkiewicz10}. Here we propose an efficient method of selecting a spanning set
of orthogonal modes based on the Gram matrix of form-factors with respect to the
appropriate scalar product defined in Eq.~\eqref{gram_matrix}.  The formal details are given in the Appendix.
	
Thus, as explained in the Appendix, one can construct the set of orthogonal modes $B_{\alpha}$ from the normalized eigenvectors $\bm{u}^{(\alpha)}=(u_{1}^{(\alpha)},u_{2}^{(\alpha)}\ldots)$ of the Gram matrix~$A$ [Eq.~\eqref{gram_matrix}]. Denoting the corresponding eigenvalues by $\lambda_{\alpha}$ we arrive at the
Hamiltonian in the form
\begin{equation}\label{HF-ortog}
H_{\mathrm{F}}=\sqrt{\frac{\hbar \Omega   e^{2}}{2l_{0}\widetilde{\epsilon}\epsilon_{0}}}  
\sum_{\alpha \nu \mu}  \sqrt{\lambda_{\alpha}}
\left( u^{(\alpha)^{*}}_{\nu \mu} B_{\alpha} c^{\dagger}_{\nu} c^{\phantom{\dagger}}_{\mu} + \mathrm{h.c.} \right ).
\end{equation}
Finally, we diagonalize this Hamiltonian in the space of zero-, one- and two-phonon states
\begin{equation*}
| \Psi_{i} \ket = \sum_{\nu} d_{\nu} | \nu \ket 
+ \sum_{\nu \alpha}  d_{ \nu \alpha} B^{\dagger}_{\alpha} | \nu \ket 
+ \sum_{\nu \alpha \beta} d_{\nu \alpha \beta} \eta_{\alpha \beta} B^{\dagger}_{\alpha} B^{\dagger}_{\beta}| \nu \ket. 
\end{equation*}
Here $\eta_{\alpha \beta}=1/\sqrt{2}$
for $\alpha = \beta$ and $1$ otherwise,
and $\nu$ corresponds to the
noninteracting electron-hole pair states defined in Sec.~\ref{sec:phonons}. As a result of this procedure, for the
exciton-polaron problem with the restricted s-shell basis (as in Fig.~\ref{fig:x_res}) the
Hamiltonian involves only several orthogonal modes out of the initial $n^{2}$ modes, where $n$ denotes number of exciton states. 

\section{Results}
\label{sec:results}

\begin{figure*}
	\begin{center}
		\includegraphics[width=460pt]{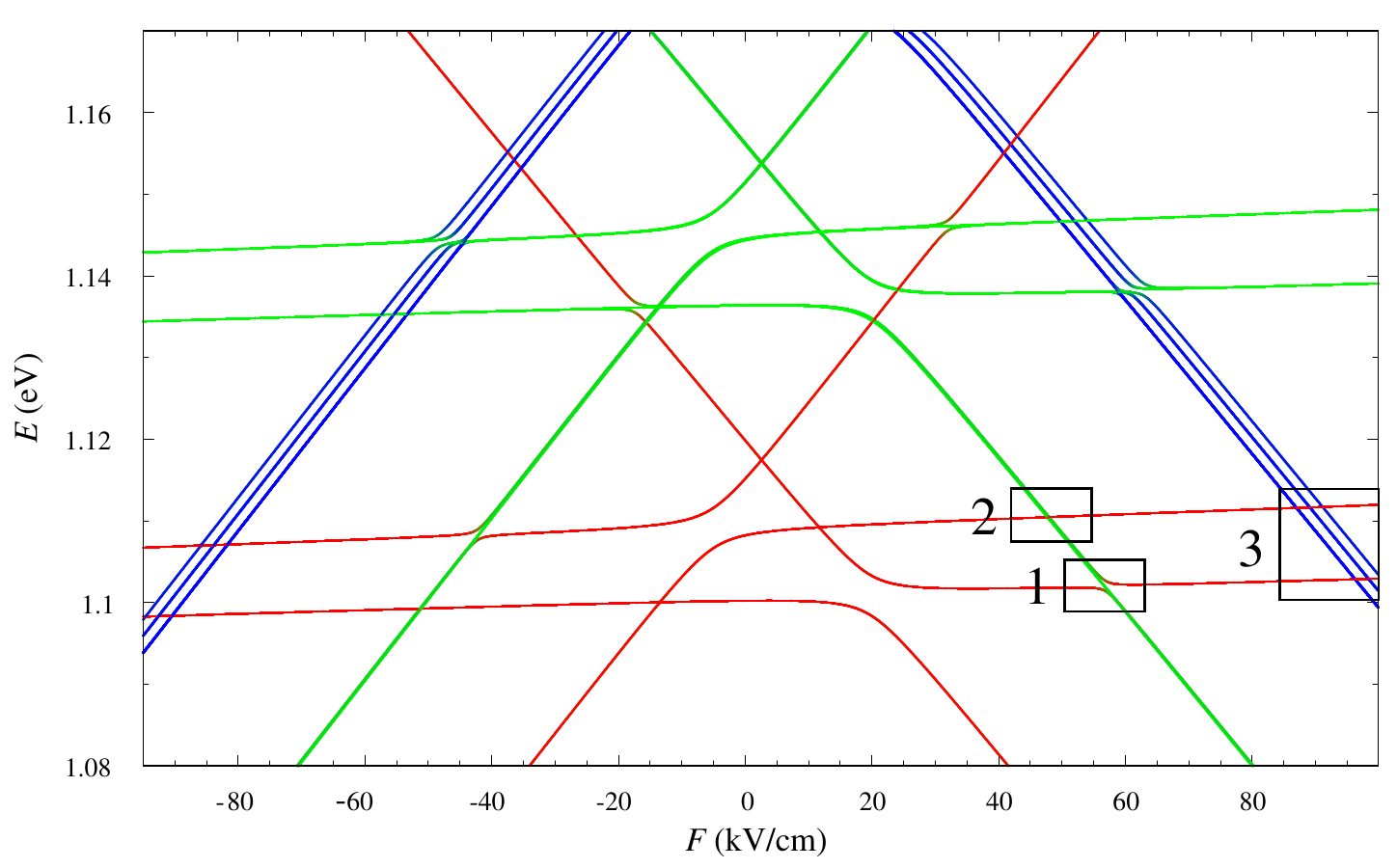}
	\end{center}
	\caption{\label{fig:pol_res}Polaron energy branches as a function of axial electric field at $D=9.6$~nm.}
\end{figure*}
The polaron energy branches were calculated as a function of an axial electric field (Fig.~\ref{fig:pol_res}).
As described in Sec.~\ref{sec:model}, every polaron state $| \Psi_{i} \ket$ is expressed
in the basis of zero-, one- and two-phonon states. We calculated the amplitudes of these
components and marked the dominating one in Fig.~\ref{fig:pol_res} by
assigning colors to the lines: red, green and blue, respectively.  The colors are mixed at the
resonances involving states with different numbers of phonons. The central part of the
plot corresponds to the states with a dominant zero-phonon (pure excitonic)
component. They show the same level structure as in Fig.~\ref{fig:x_res} with only a
small shift. The same pattern is reflected in $n$-phonon replicas at the
energies shifted by approximately $n \hbar \Omega$. Box $1$ contains an avoided level
crossing between the direct exciton state (localized in the upper dot) and the
single-phonon replica of the indirect exciton state (the hole in the upper dot and the
electron in the lower dot). Thus, this anti-crossing corresponds to the resonant
electron tunneling combined with emission/absorption of a single LO phonon (decoupled phonon modes do not lead to an anti-crossing). A similar
situation takes place for the hole (box $2$). However, for the considered DQD, its
coupling strength is much weaker than in the electron case. Furthermore, in contrast to
the electron case (which already converges in a small basis), the accurate treatment of
the hole-phonon resonances would require larger basis which rapidly increases the
computation cost.   
Therefore, we limit our present discussion to the most pronounced electron-phonon
resonances. The resonances marked by the box $3$ involve the coupling to two-phonon
states. They occur at very large electric fields and represent a second order processes
with much weaker coupling strength compared to the single-phonon case. Furthermore, a proper treatment of two-phonon states requires taking into account $3$-phonon states\cite{kaczmarkiewicz10}. 
However, neglecting
two-phonon states would result in the appearance of non-physical resonances in the
single-phonon spectra. 

 \begin{figure}
 	\begin{center}
 		\includegraphics[width=230pt]{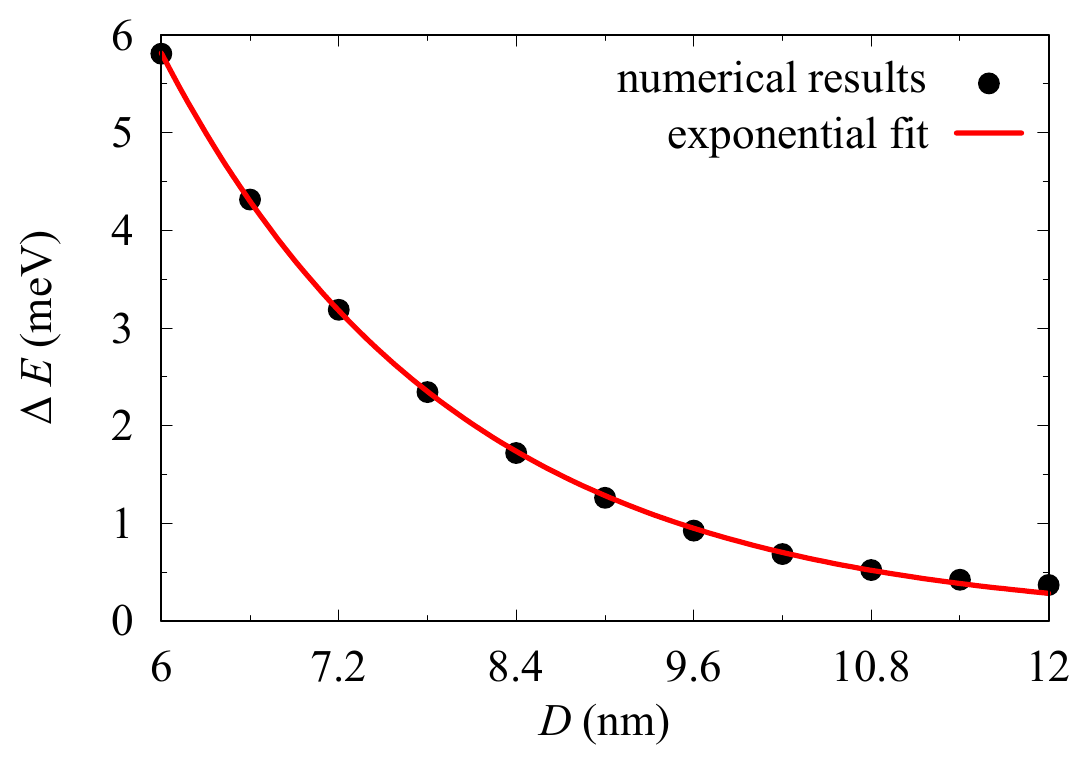}
 	\end{center}
 \caption{\label{fig:distP}The width of avoided crossing related to electron
           tunneling combined with emission/absorption of LO phonon (box 1 in Fig.~\ref{fig:pol_res}).}
 \end{figure}
An interesting question is the dependence of the tunnel coupling strength (extracted from
the numerical results as the half-width of the resonant splitting) on the inter-dot
distance. The well-known one-dimensional model of tunneling yields exponential
dependence. Such a behavior is indeed obtained for an electron in a DQD structure
\cite{gawarecki10} (which is not obvious, as demonstrated by the hole-related
counter-example \cite{jaskolski04,jaskolski06}). Therefore, we have investigated the width
of the electron-phonon resonance (corresponding to the box $1$) 
as a function of the distance between the dots. Similarly to the direct tunnel resonance, we
obtained an exponential decay (Fig.~\ref{fig:distP}) of the avoided crossing width. The
exponential fitting by the function $f(D) = a \exp(-b D)$ yields $a = 0.119$~eV and b =
$0.503$~eV/nm. Apart from the much lower amplitude, which is expected for a
phonon-assisted process, the decay rate $b$ of the polaron resonance is about $10$\% lower in
comparison to the direct resonance. This could be due to the higher energy of the
involved states, and, in consequence, to the deeper penetration into the barrier. 
	
 \begin{figure*}
 \begin{center}
 	\includegraphics[width=490pt]{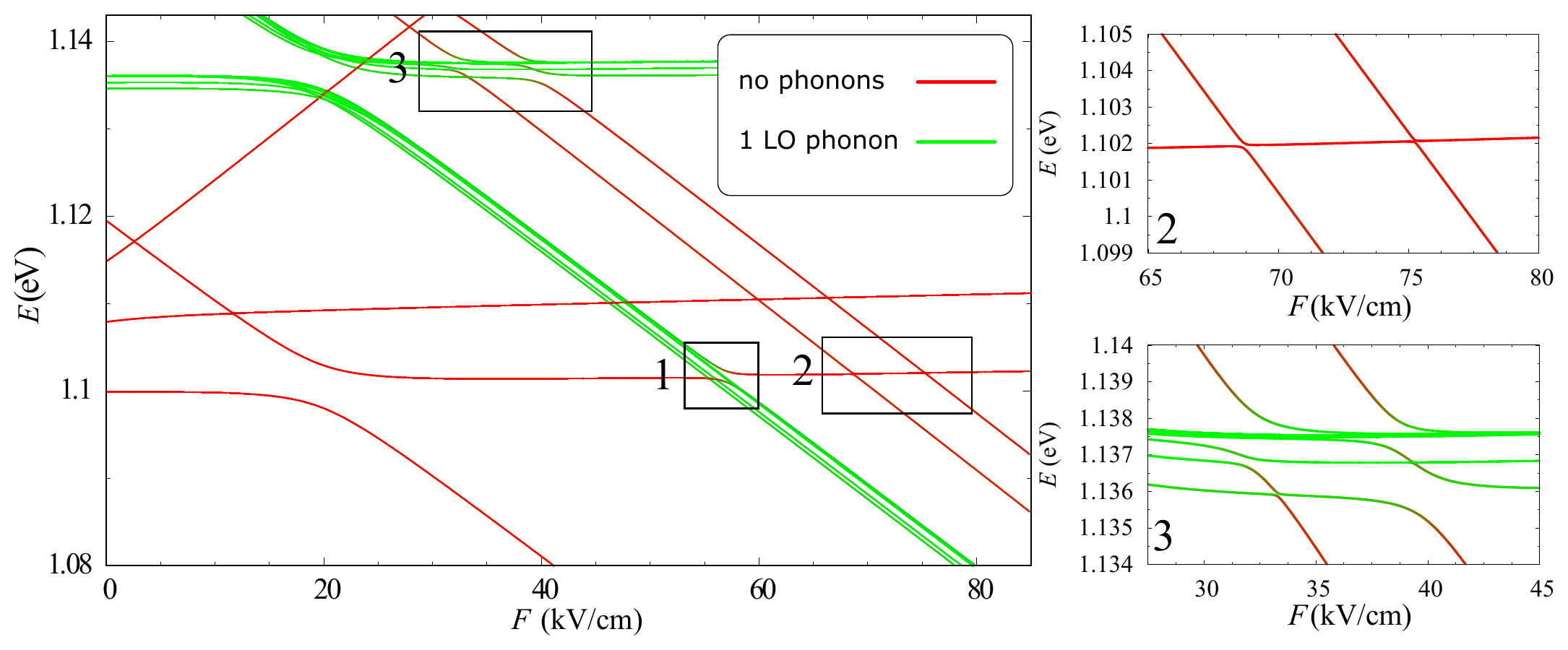}
 \end{center}
 \caption{\label{fig:pol_sp}Polaron energy branches as a function of axial
   electric field at $D=9.6$~nm. The panels on the right show magnified pictures of the corresponding regions.} 
 \end{figure*}
In the next step, we study the coupling between the electron $p$-shell states located in the upper dot and
the phonon replica of the $s$-state from the lower one. To this end, we extended the
electron basis to $12$ states ($s$, $p$ shells in each dot) while the hole basis still contains $4$ states. 
The results are shown in Fig.~\ref{fig:pol_sp}. The basis extension increases the number of orthogonal phonon modes. The exciton-phonon anti-crossing which involves $s$-states (marked in box~$1$) has a very similar width (the difference less than $1$\%) to that obtained in the reduced basis (box 1 in Fig.~\ref{fig:pol_res}).
The box $2$ contains avoided crossing related to the resonant transition between $s$
and $p$ states of the different dots. Since the angular momentum has to be conserved, this
coupling is possible only if the axial symmetry of the system is broken. This can be
caused e.g. by the relative displacement of the dots\cite{gawarecki14}, bulk inversion
asymmetry (BIA)\cite{winkler03} or emerge from the atomistic
structure\cite{bester05,zielinski15}. However, in the present model we assume perfectly aligned dots
and the only coupling mechanism is due to BIA and the interface. This leads to narrow avoided crossings
visible in box $2$. On the other hand, phonons carry angular momentum and
can couple states with different values of the angular momentum. In consequence,
we observe the pronounced anti-crossings between the electron states from $p$-shell and the phonon replicas of the $s$-shell states 
(see box $3$). This LO-phonon assisted coupling is much stronger than those resulting from
BIA (box $2$).    
	
\section{Summary}
\label{sec:conclusions}

We calculated polaron states for excitons in self-assembled double quantum
dots using realistic wave functions obtained from \kp and configuration-interaction
calculations. We proposed a mode orthogonalization scheme that yields a spanning set of
effectively coupled collective modes. We investigated resonances (avoided level crossings)
related to the electron resonant tunneling combined with emission/absorption of an LO
phonon. We have found that the strength of this LO-phonon mediated coupling shows
exponential dependence on the inter-dot distance, like in the one-dimensional tunneling
problem, with a characteristic length comparable to the direct (zero-phonon) tunneling resonance but
with a smaller amplitude. We have also shown that LO phonon modes can efficiently couple
states that belong to different shells ($s$ and $p$) from different dots. In this case,
direct resonance is strongly suppressed by angular momentum selection rules, while the
LO-phonon-assisted coupling is allowed and has a larger amplitude. 
	
\acknowledgments

P. K. acknowledges support by the Grant No. 2014/12/T/ST3/00231 from the Polish National
Science Centre (Narodowe Centrum Nauki). 
K. G. acknowledges support by the Grant No. 2012/05/N/ST3/03079 from the Polish National
Science Centre (Narodowe Centrum Nauki). 
Calculations have been carried out in Wroclaw Centre for Networking and Supercomputing
(http://www.wcss.wroc.pl), grant No. 203.

\appendix
\section{Orthogonal phonon modes}

In this appendix we present a computationally effective procedure of identifying a set of
orthogonal collective modes that span the space of all the phonon modes coupled to the
carrier subsystem. We will use the combined index $i=(\mu,\nu)$ for the form-factors and
define the scalar product 
\begin{displaymath}
\left( \ff_{i},\ff_{i'}\right) = \sumq \frac{l_{0}}{Vq^{2}}\ff_{i}^{*}(\qq)\ff_{i'}(\qq).
\end{displaymath}
In the sense of this scalar product, the matrix $A$ defined in Sec.~\ref{sec:collective} is the Gram matrix of
the functions $\ff_{i}$, $A_{ij}=( \ff_{i},\ff_{j})$. Being a Gram matrix, it is obviously
hermitian and also positive semi-definite: for any vector $\bm{u}=(u_{1},u_{2},\ldots)$, 
\begin{equation}\label{positive}
\bm{u}^{\dag}A\bm{u}=\left( \sum_{i}u_{i}\ff_{i},\sum_{i}u_{i}\ff_{i}\right) \ge 0.
\end{equation}
It is known that the Gram matrix carries information on the linear dependence of the system
of vectors. Specifically, the set of form factors $\ff_{i}$ satisfies a linear dependence relation
$\sum_{i}u_{i}\ff_{i}=0$ if and only if $\bm{u}=(u_{1},u_{2},\ldots)$ is in the kernel of
$A$. Indeed, from Eq.~\eqref{positive}, such linear dependence implies
$\bm{u}^{\dag}A\bm{u}=0$. For a positive semi-definite matrix this is only possible if
$\bm{u}$ is in the kernel. Conversely $A\bm{u}=0$ implies
$\sum_{i}(\ff_{j},\ff_{i})u_{i}=0$, hence $(\sum_{j}u_{j}\ff_{j},\sum_{i}u_{i}\ff_{i})=0$,
which means that $\sum_{i}u_{i}\ff_{i}=0$. 

This algebraic structure allows us to express the Fr\"ohlich Hamiltonian in terms of a set
of orthogonal modes at a cost of generating and diagonalizing the Gram matrix $A$, the
size of which is typically rather modest. Assume that $\bm{u}^{(\alpha)}$ are normalized
eigenvectors of $A$ with eigenvalues $\lambda_{\alpha}$, that is, $A
\bm{u}^{(\alpha)}=\lambda_{a}\bm{u}^{(\alpha)}$. Define 
$\tilde{\mathcal{G}}_{\alpha}=\sum_{i}u_{i}^{(\alpha)}\ff_{i}$. Clearly, from the statement demonstrated above, $\tilde{\mathcal{G}}_{\alpha}=0$ if
and only if $\lambda_{\alpha}=0$.   One easily shows that 
$(\tilde{\mathcal{G}}_{\alpha}, \tilde{\mathcal{G}}_{\alpha'})=
\lambda_{\alpha}\delta_{\alpha\alpha'}$. For $\alpha$ such that $\lambda_{\alpha}\neq 0$,
define the normalized mode function 
$\mathcal{G}_{\alpha}=\tilde{\mathcal{G}}_{\alpha}/\sqrt{\lambda_{\alpha}}$ and the modes 
\begin{equation}\label{B-alfa}
B_{\alpha}=\sumq \sqrt{\frac{l_{0}}{V}}\frac{1}{q}\mathcal{G}_{\alpha}(\qq)b_{\qq}.
\end{equation}
These modes are orthogonal, i.e., they satisfy the canonical commutation relations
$[B_{\alpha}, B_{\alpha'}^{\dag}]=\delta_{\alpha\alpha'}$ etc. 
On the other hand, completeness of the set
$\{\bm{u}^{(\alpha)}\}$ implies that
$\ff_{i}=\sum_{\alpha}u_{i}^{(\alpha)}\tilde{\mathcal{G}}_{\alpha}$. 
Substituting this to the Fr\"ohlich Hamiltonian [Eq.~\eqref{HF-X}] (the treatment of the
single-particle Hamiltonian is the same), one arrives at the final form of the Fr\"ohlich
Hamiltonian expressed in terms of the orthogonal modes, given in Eq.~\eqref{HF-ortog}.

\bibliographystyle{prsty}
\bibliography{abbr,quantum,PM}

\end{document}